\documentclass{article}
\usepackage{eurosym}
\usepackage{amssymb}
\usepackage{amsmath}
\usepackage{cite}
\usepackage{graphicx}
\usepackage{epsf}

\input{tcilatex}

\begin{document}

\title{Lie symmetry analysis and one-dimensional optimal system for the
generalized 2+1 Kadomtsev-Petviashvili equation}
\author{Andronikos Paliathanasis\thanks{%
Email: anpaliat@phys.uoa.gr} \\
{\ \textit{Institute of Systems Science, Durban University of Technology }}\\
{\ \textit{PO Box 1334, Durban 4000, Republic of South Africa}}}
\maketitle

\begin{abstract}
We classify the Lie point symmetries for the 2+1 nonlinear generalized
Kadomtsev-Petviashvili equation by determine all the possible $f\left(
u\right) $ functional forms where the latter depends. For each case the
one-dimensional optimal system is derived; a necessary analysis to find all
the possible similarity transformations which simplify the equation. We
demonstrate our results by constructing static and travel-wave similarity
solutions. In particular the latter solutions satisfy a second-order
nonlinear ordinary differential equation which can be solved by quadratures.%
\end{abstract}

\section{Introduction}

There are many different approaches to study nonlinear differential
equations and determine analytical solutions \cite%
{olver,meth2,meth3,meth4,meth5,meth6,meth7,meth8,meth9}. A systematic method
which has been widely applied with many interesting results was established
by S. Lie at the end of the 19th century, and it is described in his work on
the theory of transformations groups \cite{lie1,lie2,lie3}.

The main novelty of Lie's theory is that the transformations groups which
leave invariant a differential equation, can be used to simplify the given
equation. In particular, Lie symmetries are applied to the simplification
process of a differential equation by means of reduction. There are
differences in the application of Lie symmetries between ordinary
differential equations (ODEs) and partial differential equations (PDEs). For
PDEs the application of a Lie point symmetry through the so-called
similarity transformation leads to a differential equation with less
independent variables and of the same order. Oppositely, in the case of ODEs
the application of a Lie symmetry reduces the order of the given
differential equation by one \cite{olver,kumei}.

The application of the theory of transformations groups in differential
equations is not restricted to the application of the similarity
transformation. Lie symmetries can be used to determine algebraic equivalent
systems as also to provide linearization criteria for nonlinear differential
equations \cite{lin1,lin2,lin3}. In addition, Lie symmetries are applied in
order to construct conservation laws \cite{con1,con2,con3}; to determine new
solutions from old solutions \cite{ne1} and many other applications \cite%
{ibra}.

The plethora of results which can be obtained by the Lie symmetries for
nonlinear differential equations have led to the algebraic classification
problem for differential equations. The first algebraic classification
scheme was performed by L.V. Ovsiannikov in 1982, who classified all the
forms of the 1+1 nonlinear PDE $u_{t}-\left( f\left( u\right) u_{x}\right)
_{x}=0$ where the latter equation admits Lie symmetries \cite{Ovsi}. In
terms of nonlinear wave equations there are various studies on the group
properties, Ames et al. classified the Lie point symmetries for the
nonlinear differential equation $u_{tt}-\left( f\left( u\right) u_{x}\right)
_{x}=0$. Applications of Lie symmetries in Shallow-water equations are
presented in \cite{sw1,sw2,sw3,sw4,sw5,sw6,sw7,sw8}; while applications of
other subjects of applied mathematics and mathematical physics are presented
in \cite{ref1,ref2,ref3,ref4,ref5,ref6,ref7,ref8,ref9,ref10,ref11,ref12} and
references therein.

In this work we focus on the algebraic classification problem for the 2+1
nonlinear generalized Kadomtsev-Petviashvili (KP) equation \cite{eq1} 
\begin{eqnarray}
u_{t}+f\left( u\right) u_{x}+u_{xxx}+\varepsilon v_{y} &=&0,  \label{kp.01}
\\
v_{x}-u_{y} &=&0,  \label{kp.02}
\end{eqnarray}%
or equivalent%
\begin{equation}
\left( u_{t}+f\left( u\right) u_{x}+u_{xxx}\right) _{x}+\varepsilon u_{yy}=0,
\end{equation}%
where $f\left( u\right) $ is an arbitrary nonlinear function, $u=u\left(
t,x,y\right) ,~v=v\left( t,x,y\right) $, while parameter $\varepsilon $ can
be normalized to $\varepsilon =\pm 1$ and it measures the traverse
dispersion effects on weakly nonlinear waves.

KP equation is recovered for the linear function $f\left( u\right) $ and it
can be seen as the extension of the Korteveg-de Vreis equation in higher
dimensions.\ Nowadays KP equation is the standard model for the description
of weakly nonlinear waves of small amplitude in various physical situations 
\cite{kp1,kp2,kp3}. The KP equation is a well-known integrable equation
which has been used as a source of integrable equations, for more details
see \cite{ss1}.

In \cite{kps01} it was found that the KP equation can be reduced to into the
Painlev\'{e} transcendental equation of the first kind by using the Lie
invariants. The Lie symmetries and the possible reductions of the KP
equations were studied also by S.-Y. Lou in \cite{kps1}; while recently the
Lie point symmetries of the KP equation with time-dependent coefficients
have been determined in \cite{kps2}, a similar analysis with and time- and
space- dependent coefficients was performed in \cite{kps3}. For other
integrable hierarchies of PDEs we refer the reader in \cite{book001}

In the following Sections we shall determine the forms of the unknown
nonlinear function $f\left( u\right) $ where the 2+1 nonlinear generalized
KP equation (\ref{kp.01}), (\ref{kp.02}) admits Lie point symmetries. For
the different functions $f\left( u\right) $ we determine the one-dimensional
optimal system of the admitted Lie point symmetries by the generalized KP
equation. The determination of the optimal system is necessary in order to
understand the possible reductions of the differential equation.

For the one-dimensional system we calculate the corresponding invariants
which define the similarity transformations to reduce the differential
equation. The results are presented in a tabular list. \ Moreover, we shall
present two examples where we show how to apply the Lie invariants and
determine similarity transformations. We shall see that for the arbitrary
functional form of $f\left( u\right) $ for the static solution and the
travel-wave solution the generalized KP equation (\ref{kp.01}), (\ref{kp.02}%
) can be solved by quadratures. While for some specific functional forms of $%
f\left( u\right) $ the solution of the original system is described by
well-known one-dimensional Newtonian systems such is the Ermakov-Pinney
equation. The outline of the paper follows.

In Section \ref{sec2}, we present the main results of our analysis, where we
determine the Lie point symmetries for the 2+1 nonlinear generalized KP
equation (\ref{kp.01}), (\ref{kp.02}) for specific forms of $f\left(
u\right) $. In particular we determine the Lie point symmetries for
arbitrary function $f\left( u\right) $, where additional symmetries exist
when $f\left( u\right) =u^{k}+f_{0}$ and $f\left( u\right) =e^{\sigma
u}+f_{0}$. For each of the cases, the one-dimensional optimal system is
calculated. In Section \ref{sec4}, we determine the Lie invariants for all
the one-dimensional systems. This invariants can be used to find similarity
transformations in order to the generalized KP equation and construct
similarity solutions. The similarity transformations are applied to find
static similarity solutions or travel-wave solutions. In Appendices \ref{ap1}
and \ref{ap2} we present the basic properties and definitions for the Lie
theory and the one-dimensional optimal system, while in Appendix \ref{ap3}
we extend our analysis and we present the Lie point symmetries for the 3+1
nonlinear generalized KP equation \cite{eq1}. Finally in Section \ref{sec5},
we discuss our results and we draw our conclusions.

\section{Classification of Lie symmetries}

\label{sec2}

In this section we solve the algebraic classification problem for the 2+1
nonlinear general KP equation of our consideration by finding all the
nonlinear functions $f\left( u\right) $ in which equations (\ref{kp.01}), (%
\ref{kp.02}) admit Lie point symmetries. In each case the one-dimensional
optimal system is derived. The Lie theory and the definition of the
one-dimensional optimal system are presented in Appendices \ref{ap1} and \ref%
{ap2} respectively.

\subsection{Arbitrary function $f\left( u\right) $}

For the arbitrary function $f\left( u\right) $ the 2+1 generalized KP
equations (\ref{kp.01}), (\ref{kp.02}) admit the following Lie point
symmetries 
\begin{equation}
X_{1}=\partial _{t}~,~X_{2}=\partial _{x}~,~X_{3}=\partial
_{y}~,~X_{4}=2\varepsilon t\partial _{y}-y\partial _{x}+u\partial
_{v}~,~X_{\beta }=\beta \left( t\right) \partial _{v}.
\end{equation}%
where function $\beta \left( t\right) $ is arbitrary.

The symmetry vector $X_{\beta }$ indicates that there are infinity number of
solutions of the form $v\left( t,x,y\right) =v\left( t\right) $ which solves
the KP equation. However it does not play any role in the determination of
the exact solutions, hence we shall omit it.

\begin{table}[tbp] \centering%
\caption{Commutators of the admitted Lie point symmetries for the 2+1
nonlinear KP equation for arbitrary function $f(u)$}%
\begin{tabular}{ccccc}
\hline\hline
$\left[ ~,~\right] $ & $\mathbf{X}_{1}$ & $\mathbf{X}_{2}$ & $\mathbf{X}_{3}$
& $\mathbf{X}_{4}$ \\ \hline
$\mathbf{X}_{1}$ & $0$ & $0$ & $0$ & $2\varepsilon X_{3}$ \\ 
$\mathbf{X}_{2}$ & $0$ & $0$ & $0$ & $0$ \\ 
$\mathbf{X}_{3}$ & $0$ & $0$ & $0$ & $-X_{2}$ \\ 
$\mathbf{X}_{4}$ & $-2\varepsilon X_{3}$ & $0$ & $X_{2}$ & $0$ \\ 
\hline\hline
\end{tabular}%
\label{tab1}%
\end{table}%

As far as the rest of the symmetry vectors are concerned, i.e. the vector
fields $X_{1},~X_{2},~X_{3}$ and $X_{4},$ we calculate the commutators which
are presented in Table \ref{tab1}. The admitted Lie algebra is the $A_{4,3}$ 
$\ $in the Morozov-Mubarakzyanov classification scheme \cite{B21,B22,B23,B24}%
, for more details we refer the reader in the review article \cite{B25}.

\subsubsection{One-dimensional optimal system}

In order to determine the one-dimensional optimal system, the adjoint
representation and the invariants of the adjoint action should be
determined. The adjoint representation of the symmetry vectors $\left\{
X_{1},X_{2},X_{3},X_{4}\right\} $ is presented in Table \ref{tab1a}.

The invariants $\phi \left( a_{i}\right) ~$of the adjoint action are
determined by the set of differential equations%
\begin{equation}
\Delta _{i}\left( \phi \right) =C_{ij}^{k}a^{j}\frac{\partial }{\partial
a^{k}}\phi ,  \label{kp.03}
\end{equation}%
where $C_{ij}^{k}$ are the structure constants of the Lie algebra.

Therefore, from (\ref{kp.03}) and Table \ref{tab1} we end up with the system
of first-order partial differential equations%
\begin{equation}
2\varepsilon a_{4}\frac{\partial \phi }{\partial a_{3}}=0~,~-a_{4}\frac{%
\partial \phi }{\partial a_{2}}=0,  \label{kp.04}
\end{equation}%
from where we infer $\phi =\phi \left( a_{1},a_{4}\right) $, that is, the
invariants of the adjoint action are the $a_{1}$ and $a_{4}$.

\begin{table}[tbp] \centering%
\caption{Adjoint representation of the admitted Lie point symmetries for the
2+1 nonlinear KP equation for arbitrary function $f(u)$}%
\begin{tabular}{ccccc}
\hline\hline
$Ad\left( \exp \left( \varepsilon X_{i}\right) \right) X_{j}$ & $\mathbf{X}%
_{1}$ & $\mathbf{X}_{2}$ & $\mathbf{X}_{3}$ & $\mathbf{X}_{4}$ \\ \hline
$\mathbf{X}_{1}$ & $X_{1}$ & $X_{2}$ & $X_{3}$ & $-2\varepsilon
^{2}X_{3}+X_{4}$ \\ 
$\mathbf{X}_{2}$ & $X_{1}$ & $X_{2}$ & $X_{3}$ & $X_{4}$ \\ 
$\mathbf{X}_{3}$ & $X_{1}$ & $X_{2}$ & $X_{3}$ & $\varepsilon X_{2}+X_{4}$
\\ 
$\mathbf{X}_{4}$ & $X_{1}-\varepsilon ^{3}X_{2}+2\varepsilon ^{2}X_{3}$ & $%
X_{2}$ & $-\varepsilon X_{2}+X_{3}$ & $X_{4}$ \\ \hline\hline
\end{tabular}%
\label{tab1a}%
\end{table}%

We define the generic symmetry vector 
\begin{equation}
X=a_{1}X_{1}+a_{2}X_{2}+a_{3}X_{3}+a_{4}X_{4},
\end{equation}%
and with the use of Table \ref{tab1a} and of the invariants of the adjoint
representation as given in Table \ref{tab1a} we have the following possible
cases

Case 1: $a_{1}=0,~a_{2}=0$. The generic symmetry vector is 
\begin{equation}
X^{\prime }=a_{2}X_{2}+a_{3}X_{3},
\end{equation}%
which gives the one-dimensional optimal system%
\begin{equation*}
\left\{ X_{2}\right\} ~,~\left\{ X_{3}\right\} ~,~\left\{ X_{2}+\gamma
X_{3}\right\} .
\end{equation*}

Case 2: $a_{1}\neq 0,~a_{2}=0$. The generic symmetry vector is%
\begin{equation}
X^{\prime \prime }=a_{1}X_{1}+a_{2}X_{2}+a_{3}X_{3},
\end{equation}%
from where we infer the additional one-dimensional algebras%
\begin{equation*}
\left\{ X_{1}\right\} ~,~\left\{ X_{1}+\gamma X_{2}\right\} ~,~\left\{
X_{1}+\delta X_{3}\right\} ~,~\left\{ X_{1}+\gamma X_{2}+\delta
X_{3}\right\} .
\end{equation*}

Case 3: $a_{1}=0,~a_{2}\neq 0.$ The generic symmetry vector is%
\begin{equation}
X^{\prime \prime \prime }=a_{2}X_{2}+a_{3}X_{3}+a_{4}X_{4},
\end{equation}%
where now the additional one-dimensional algebras are%
\begin{equation*}
\left\{ X_{4}\right\} ~,~\left\{ X_{4}+\gamma X_{2}\right\} ~,~\left\{
X_{4}+\delta X_{3}\right\} .
\end{equation*}

Case 4: $a_{1}a_{2}\neq 0$. In the generic case the additional
one-dimensional Lie algebra is found to be 
\begin{equation*}
\left\{ X_{1}+\gamma X_{4}\right\} .
\end{equation*}

Hence, the one-dimensional optimal system for the 2+1 generalized KP
equation (\ref{kp.01}), (\ref{kp.02}) for arbitrary function $f\left(
u\right) $ consists by the Lie algebras%
\begin{eqnarray*}
&&\left\{ X_{1}\right\} ~,~\left\{ X_{2}\right\} ~,~\left\{ X_{3}\right\}
~,~\left\{ X_{4}\right\} ~,~\left\{ X_{2}+\gamma X_{3}\right\} ~,~\left\{
X_{1}+\gamma X_{2}\right\} ~,~\left\{ X_{1}+\delta X_{3}\right\} ~, \\
&&\left\{ X_{1}+\gamma X_{2}+\delta X_{3}\right\} ,~\left\{ X_{4}+\gamma
X_{2}\right\} ~,~\left\{ X_{4}+\delta X_{3}\right\} ~,~\left\{ X_{1}+\gamma
X_{4}\right\} .
\end{eqnarray*}

\subsection{Power-law $f\left( u\right) =u^{k}+f_{0}$}

When $f\left( u\right) $ is a power law function, that is, $f\left( u\right)
=u^{k}+f_{0}$ the admitted Lie symmetries for equation (\ref{kp.01}), (\ref%
{kp.02}) are 
\begin{eqnarray}
X_{1} &=&\partial _{t}~,~X_{2}=\partial _{x}~,~X_{3}=\partial
_{y}~,~X_{4}=2\varepsilon t\partial _{y}-y\partial _{x}+u\partial _{v}~,~ 
\notag \\
X_{5} &=&2u\partial _{u}+\left( k+2\right) v\partial _{v}-k\left( 3t\partial
_{t}+\left( x+2f_{0}t\right) \partial _{x}+2y\partial _{y}\right)
~,~X_{\beta }=\beta \left( t\right) \partial _{v},~
\end{eqnarray}%
where again $\beta \left( t\right) $ is an arbitrary function and $X_{5}$ is
an extra Lie point symmetry. We observe that $X_{5}$ is a scaling symmetry.
The commutators of the admitted Lie point symmetries are given in Table \ref%
{tab2}. The admitted Lie point symmetries form the $A_{5,37}$ Lie algebra in
the Patera et al. classification scheme \cite{B26}.

\begin{table}[tbp] \centering%
\caption{Commutators of the admitted Lie point symmetries for the 2+1
nonlinear KP equation for power-law function $f(u)$}%
\begin{tabular}{cccccc}
\hline\hline
$\left[ ~,~\right] $ & $\mathbf{X}_{1}$ & $\mathbf{X}_{2}$ & $\mathbf{X}_{3}$
& $\mathbf{X}_{4}$ & $\mathbf{X}_{5}$ \\ \hline
$\mathbf{X}_{1}$ & $0$ & $0$ & $0$ & $2\varepsilon X_{3}$ & $%
-3kX_{1}-2kf_{0}X_{2}$ \\ 
$\mathbf{X}_{2}$ & $0$ & $0$ & $0$ & $0$ & $-kX_{2}$ \\ 
$\mathbf{X}_{3}$ & $0$ & $0$ & $0$ & $-X_{2}$ & $-2kX_{3}$ \\ 
$\mathbf{X}_{4}$ & $-2\varepsilon X_{3}$ & $0$ & $X_{2}$ & $0$ & $kX_{4}$ \\ 
$\mathbf{X}_{5}$ & $3kX_{1}+2kf_{0}X_{2}$ & $kX_{2}$ & $2kX_{3}$ & $-kX_{4}$
& $0$ \\ \hline\hline
\end{tabular}%
\label{tab2}%
\end{table}%

\subsubsection{One-dimensional optimal system}

\begin{table}[tbp] \centering%
\caption{Adjoint representation of the admitted Lie point symmetries for the
2+1 nonlinear KP equation for power-law function $f(u)$}%
\begin{tabular}{cccccc}
\hline\hline
$Ad\left( \exp \left( \varepsilon X_{i}\right) \right) X_{j}$ & $\mathbf{X}%
_{1}$ & $\mathbf{X}_{2}$ & $\mathbf{X}_{3}$ & $\mathbf{X}_{4}$ & $\mathbf{X}%
_{5}$ \\ \hline
$\mathbf{X}_{1}$ & $X_{1}$ & $X_{2}$ & $X_{3}$ & $-2\varepsilon
^{2}X_{3}+X_{4}$ & $3\varepsilon kX_{1}+2\varepsilon kf_{0}X_{2}+X_{5}$ \\ 
$\mathbf{X}_{2}$ & $X_{1}$ & $X_{2}$ & $X_{3}$ & $X_{4}$ & $\varepsilon
kX_{2}+X_{5}$ \\ 
$\mathbf{X}_{3}$ & $X_{1}$ & $X_{2}$ & $X_{3}$ & $\varepsilon X_{2}+X_{4}$ & 
$2\varepsilon kX_{3}+X_{5}$ \\ 
$\mathbf{X}_{4}$ & $X_{1}-\varepsilon ^{3}X_{2}+2\varepsilon ^{2}X_{3}$ & $%
X_{2}$ & $-\varepsilon X_{2}+X_{3}$ & $X_{4}$ & $-\varepsilon kX_{4}+X_{5}$
\\ 
$\mathbf{X}_{5}$ & $e^{-3k\varepsilon }X_{1}+e^{-k\varepsilon }f_{0}\left(
e^{-2k\varepsilon }-1\right) X_{2}$ & $e^{-k\varepsilon }X_{2}$ & $%
e^{-2k\varepsilon }X_{3}$ & $e^{k\varepsilon }X_{4}$ & $X_{5}$ \\ 
\hline\hline
\end{tabular}%
\label{tab2a}%
\end{table}%

The invariants of the adjoint action are determined by the system of
first-order differential equations%
\begin{eqnarray}
2\varepsilon a_{4}\frac{\partial \phi }{\partial a_{3}}-a_{5}k\left( 3\frac{%
\partial \phi }{\partial a_{1}}+2f_{0}\frac{\partial \phi }{\partial a_{2}}%
\right) &=&0, \\
k\frac{\partial \phi }{\partial a_{2}} &=&0, \\
a_{4}\frac{\partial \phi }{\partial a_{2}}+2a_{4}k\frac{\partial \phi }{%
\partial a_{3}} &=&0, \\
-2\varepsilon a_{1}\frac{\partial \phi }{\partial a_{3}}+a_{3}\frac{\partial
\phi }{\partial a_{2}}+ka_{5}\frac{\partial \phi }{\partial a_{4}} &=&0.
\end{eqnarray}%
The latter system provides that $\phi =\phi \left( a_{5}\right) $, which
means that $a_{5}$ is the unique invariant.

Indeed when $a_{5}=0$ we find the one-dimensional optimal system of the case
where $f\left( u\right) $ is arbitrary. However, for $a_{5}\neq 0$ the
additional one-dimensional algebra is found to be the $\left\{ X_{5}\right\} 
$.

In order to demonstrate it, let us consider the generic symmetry vector%
\begin{equation}
Y=a_{1}X_{1}+a_{2}X_{2}+a_{3}X_{3}+a_{4}X_{4}+a_{5}X_{5},
\end{equation}%
then by using the he adjoint representation of the symmetry vectors $\left\{
X_{1},X_{2},X_{3},X_{4},X_{5}\right\} $ as presented in Table \ref{tab2a} we
find 
\begin{eqnarray}
Y^{\prime } &=&Ad\left( \exp \left( \varepsilon _{4}X_{4}\right) \right) Y 
\notag \\
&=&a_{1}X_{1}+\left( -\varepsilon _{4}^{3}-\varepsilon _{4}a_{3}\right)
X_{2}+\left( a_{3}+2\varepsilon _{4}^{2}\right) X_{3}+\left(
a_{4}-a_{5}\varepsilon _{4}k\right) X_{4}+a_{5}X_{5},
\end{eqnarray}%
where for $a_{5}k\varepsilon =a_{4}$ it becomes%
\begin{equation}
Y^{\prime }=a_{1}X_{1}+a_{2}^{\prime }X_{2}+a_{3}^{^{\prime
}}X_{3}+a_{5}X_{5}.
\end{equation}%
We continue by considered the adjoint transformation%
\begin{equation}
Y^{\prime \prime }=Ad\left( \exp \left( \varepsilon _{3}X_{3}\right) \right)
Y^{\prime }=a_{1}X_{1}+\left( a_{2}^{\prime }+a_{3}^{^{\prime }}\varepsilon
_{3}\right) X_{2}+\left( a_{3}^{^{\prime }}+2a_{5}\varepsilon _{4}k\right)
X_{3}+a_{5}X_{5},
\end{equation}%
and for $2a_{5}\varepsilon _{4}k=-a_{3}^{^{\prime }}$ it becomes%
\begin{equation}
Y^{\prime \prime }=Ad\left( \exp \left( \varepsilon _{3}X_{3}\right) \right)
Y^{\prime }=a_{1}X_{1}+a_{2}^{\prime \prime }X_{2}+X_{3}+a_{5}X_{5}.
\end{equation}

In addition we find 
\begin{equation}
Y^{\prime \prime \prime }=Ad\left( \exp \left( \varepsilon _{1}X_{1}\right)
\right) Y^{\prime \prime }=a_{2}^{^{\prime \prime \prime
}}X_{2}+a_{5}X_{5}~,~\text{with }a_{1}=-a_{5}3\varepsilon k,
\end{equation}%
and finally%
\begin{equation}
Y^{\prime \prime \prime \prime }=Ad\left( \exp \left( \varepsilon
_{1}X_{1}\right) \right) Y^{\prime \prime \prime }=a_{5}X_{5}~,~\alpha
_{2}^{^{\prime \prime \prime }}=-a_{5}\varepsilon k.
\end{equation}

\subsection{Exponential $f\left( u\right) =e^{\protect\sigma u}+f_{0}$}

The last case where $f\left( u\right) $ is an exponential function, that is, 
$f\left( u\right) =e^{\sigma u}+f_{0}$, the admitted Lie point symmetries by
equation (\ref{kp.01}), (\ref{kp.02}) \ are 
\begin{eqnarray}
X_{1} &=&\partial _{t}~,~X_{2}=\partial _{x}~,~X_{3}=\partial
_{y}~,~X_{4}=2\varepsilon t\partial _{y}-y\partial _{x}+u\partial _{v}~,~ 
\notag \\
\bar{X}_{5} &=&2\partial _{u}+\sigma v\partial _{v}-\sigma \left( 3t\partial
_{t}+\left( x+2f_{0}t\right) \partial _{x}+2y\partial _{y}\right)
~,~X_{6}=\partial _{v}~,~X_{\beta }=\beta \left( t\right) \partial _{v},
\end{eqnarray}%
where $\beta \left( t\right) $ is an arbitrary function. Remark that the
additional Lie point symmetry is the $\bar{X}_{5}$ while the symmetry vector 
$X_{6}$ is included into the infinity number of symmetries $X_{\beta }$.
However, in this case it is important to consider it separately in order to
define the closed algebra of the symmetry vectors $\left\{
X_{1},X_{2},X_{3},X_{4},\bar{X}_{5},X_{6}\right\} $. From the commutator of
Table \ref{tab3} we infer that the six Lie symmetries form the Lie algebra $%
\left\{ A_{5,37}\otimes _{s}A_{1}\right\} $,~$\ $where $\otimes _{s}$
denotes semi-direct product of the two Lie algebras, namely $A_{5,37}$ and $%
A_{1}$, see for details \cite{B26}.

\begin{table}[tbp] \centering%
\caption{Commutators of the admitted Lie point symmetries for the 2+1
nonlinear KP equation for exponential function $f(u)$}%
\begin{tabular}{ccccccc}
\hline\hline
$\left[ ~,~\right] $ & $\mathbf{X}_{1}$ & $\mathbf{X}_{2}$ & $\mathbf{X}_{3}$
& $\mathbf{X}_{4}$ & $\mathbf{\bar{X}}_{5}$ & $\mathbf{X}_{6}$ \\ \hline
$\mathbf{X}_{1}$ & $0$ & $0$ & $0$ & $2\varepsilon X_{3}$ & $-3\sigma
X_{1}-2\sigma f_{0}X_{2}$ & $0$ \\ 
$\mathbf{X}_{2}$ & $0$ & $0$ & $0$ & $0$ & $-\sigma X_{2}$ & $0$ \\ 
$\mathbf{X}_{3}$ & $0$ & $0$ & $0$ & $-X_{2}$ & $-2\sigma X_{3}$ & $0$ \\ 
$\mathbf{X}_{4}$ & $-2\varepsilon X_{3}$ & $0$ & $X_{2}$ & $0$ & $\sigma
X_{4}-2X_{6}$ & $0$ \\ 
$\mathbf{\bar{X}}_{5}$ & $3\sigma X_{1}+2\sigma f_{0}X_{2}$ & $\sigma X_{2}$
& $2\sigma X_{3}$ & $-\sigma X_{4}+2X_{6}$ & $0$ & $\sigma X_{6}$ \\ 
$\mathbf{X}_{6}$ & $0$ & $0$ & $0$ & $-\sigma X_{6}$ & $0$ & $0$ \\ 
\hline\hline
\end{tabular}%
\label{tab3}%
\end{table}%

\subsubsection{One-dimensional optimal system}

In order to find the one-dimensional optimal system for the case where $%
f\left( u\right) $ is an exponential function. To do that we need the
Adjoint representation which is presented in Tables \ref{tab3a} and \ref%
{tab3b}. We apply the same procedure as before, for the power-law potential
from where we find that the additional one-dimensional algebras is again the
vector field $\left\{ \bar{X}_{5}\right\} $.

The question which is raised, is about the one-dimensional optimal system
when the infinity number of symmetries, i.e. $X_{\beta }$, is included.
Recall that we should reduce the equation first from a partial differential
equation into an ordinary differential equation and the application of $%
X_{\beta }$ does not perform such process. For that reason we have not
included it in the presentation.

We continue our analysis by applying the Lie point symmetries in order to
determine the similarity transformations and when it is feasible and to
specify similarity solutions.

\begin{table}[tbp] \centering%
\caption{Adjoint representation of the admitted Lie point symmetries for the
2+1 nonlinear KP equation for exponential function $f(u)$}%
\begin{tabular}{cccc}
\hline\hline
$Ad\left( \exp \left( \varepsilon X_{i}\right) \right) X_{j}$ & $X_{1}$ & $%
X_{2}$ & $X_{3}$ \\ \hline
$X_{1}$ & $X_{1}$ & $X_{2}$ & $X_{3}$ \\ 
$X_{2}$ & $X_{1}$ & $X_{2}$ & $X_{3}$ \\ 
$X_{3}$ & $X_{1}$ & $X_{2}$ & $X_{3}$ \\ 
$X_{4}$ & $X_{1}-\varepsilon ^{3}X_{2}+2\varepsilon ^{2}X_{3}$ & $X_{2}$ & $%
-\varepsilon X_{2}+X_{3}$ \\ 
$\bar{X}_{5}$ & $e^{-3\sigma \varepsilon }X_{1}+f_{0}e^{-\sigma \varepsilon
}\left( e^{-2\sigma \varepsilon }-1\right) X_{2}$ & $e^{-\sigma \varepsilon
}X_{2}$ & $e^{-2\sigma \varepsilon }X_{3}$ \\ 
$X_{6}$ & $X_{1}$ & $X_{2}$ & $X_{3}$ \\ \hline\hline
\end{tabular}%
\label{tab3a}%
\end{table}
%

\begin{table}[tbp] \centering%
\caption{Adjoint representation of the admitted Lie point symmetries for the
2+1 nonlinear KP equation for exponential function $f(u)$}%
\begin{tabular}{cccc}
\hline\hline
$Ad\left( \exp \left( \varepsilon X_{i}\right) \right) X_{j}$ & $X_{4}$ & $%
\bar{X}_{5}$ & $X_{6}$ \\ \hline
$X_{1}$ & $-2\varepsilon ^{2}X_{3}+X_{4}$ & $\sigma \varepsilon \left(
3X_{1}+2f_{0}X_{2}\right) +\bar{X}_{5}$ & $X_{6}$ \\ 
$X_{2}$ & $X_{4}$ & $\sigma \varepsilon X_{2}+\bar{X}_{5}$ & $X_{6}$ \\ 
$X_{3}$ & $\varepsilon X_{2}+X_{4}$ & $2\sigma \varepsilon X_{3}+\bar{X}_{5}$
& $X_{6}$ \\ 
$X_{4}$ & $X_{4}$ & $-\sigma \varepsilon X_{4}+X_{5}+2\varepsilon X_{6}$ & $%
X_{6}$ \\ 
$\bar{X}_{5}$ & $e^{\sigma \varepsilon }X_{4}-2\varepsilon e^{\sigma
\varepsilon }X_{6}$ & $X_{5}$ & $e^{\sigma \varepsilon }X_{6}$ \\ 
$X_{6}$ & $X_{4}$ & $X_{5}-\varepsilon \sigma X_{6}$ & $X_{6}$ \\ 
\hline\hline
\end{tabular}%
\label{tab3b}%
\end{table}%

\section{Similarity transformations}

\label{sec4}

The main application of the Lie symmetries is that similarity
transformations can be defined which can be used to simplify the
differential equation. As far as partial differential equations are
concerned the similarity transformations are applied to reduce the number of
indepedent variables. On the contrary, in the case of ordinary differential
equations the application of similarity transformations lead to a
differential equation of lower-order. In the ideal scenario, where the
admitted Lie point symmetries are sufficient to reduce a partial
differential equation into an ordinary differential equation and the latter
equation into an algebraic equation, or into another well-known integrable
equation, with well-known solutions; we shall say that we have found a
similarity solution for the original problem.

However, the application of a similarity transformation to a given
differential equation leads to a new differential equation where it has
different algebraic properties, that is, it admits different Lie symmetries.
There is a criterion in which the Lie point symmetries of the original
equation are also point symmetries of the reduced equation. Consider the Lie
point symmetries $X_{1},$ $X_{2}$ with commutator $[X_{1},X_{2}]=cX_{2}$
where $c$ may be zero.\ Then reduction by $X_{1}$ ~in the original equation
results that $X_{2}$ being a nonlocal symmetry for the reduced equation;
while reduction by $X_{2}$ results in $X_{1}$ being an inherited Lie
symmetry of the reduced differential equation \cite{l1}. It is possible the
reduced equation to admit extra Lie point symmetries, these are called
hidden symmetries and can be used to perform further reduction \cite%
{Govinder2001}.

Before we proceed with the application of the Lie symmetries to determine
similarity solutions for the 2+1 nonlinear generalized KP equation, we
calculate the Lie invariants which correspond to all the above
one-dimensional Lie algebras. The Lie invariants are presented in Table \ref%
{tab4}.

\bigskip 

\begin{table}[tbp] \centering%
\caption{Lie invariants for the optimal system of the 2+1 nonlinear
generalized KP equation}%
\begin{tabular}{cc}
\hline\hline
\textbf{Symmetry} & \textbf{\ Invariants} \\ 
$\mathbf{X}_{1}$ & $x,y,u\left( x,y\right) ,v\left( x,y\right) $ \\ 
$\mathbf{X}_{2}$ & $t,y,u\left( t,y\right) ,v\left( t,y\right) $ \\ 
$\mathbf{X}_{3}$ & $t,x,u\left( t,x\right) ,v\left( t,x\right) $ \\ 
$\mathbf{X}_{4}$ & $t,4\varepsilon tx+y^{2},U\left( t,4\varepsilon
tx+y^{2}\right) ,~V\left( t,4\varepsilon tx+y^{2}\right) +\frac{1}{%
2\varepsilon t}U\left( t,4\varepsilon tx+y^{2}\right) $ \\ 
$\mathbf{X}_{2}+\gamma \mathbf{X}_{3}$ & $t~,~y-\gamma x,~u\left( t,y-\gamma
x\right) ,~v\left( t,y-\gamma x\right) $ \\ 
$\mathbf{X}_{1}+\gamma \mathbf{X}_{2}$ & $y~,x-\gamma t,~u\left( y,x-\gamma
t\right) ,~v\left( y,x-\gamma t\right) $ \\ 
$\mathbf{X}_{1}+\gamma \mathbf{X}_{3}$ & $x~,y-\gamma t,~u\left( x,y-\gamma
t\right) ,~v\left( x,y-\gamma t\right) $ \\ 
$\mathbf{X}_{1}+\gamma \mathbf{X}_{2}+\delta \mathbf{X}_{3}$ & $x-\gamma
t,y-\delta t,u\left( x-\gamma t,y-\delta t\right) ~,~v\left( x-\gamma
t,y-\delta t\right) $ \\ 
$\mathbf{X}_{4}+\gamma \mathbf{X}_{2}$ & $t,\zeta =\frac{2\gamma
x-x^{2}-4\varepsilon ty}{4\varepsilon t},U\left( t,\zeta \right) ,V\left(
t,\zeta \right) +\frac{\left( \gamma -x\right) }{2\varepsilon t}U\left(
t,\zeta \right) $ \\ 
$\mathbf{X}_{4}+\gamma \mathbf{X}_{3}$ & $t,\omega =\frac{2\gamma
y-y^{2}-4\varepsilon tx}{4\varepsilon t},U\left( t,\omega \right) ,V\left(
t,\omega \right) +\frac{\left( \gamma -y\right) }{2\varepsilon t}U\left(
t,\omega \right) $ \\ 
$\mathbf{X}_{1}+\gamma \mathbf{X}_{4}$ & $\xi =y-\varepsilon \gamma
t^{2},\zeta =x-\frac{2\gamma ^{2}}{3}\varepsilon t^{3}+\gamma yt,U\left( \xi
,\zeta \right) ,V\left( \xi ,\zeta \right) +\gamma U\left( \xi ,\zeta
\right) $ \\ 
$\mathbf{\bar{X}}_{5}$ & $\left( x-f_{0}t\right) t^{-\frac{1}{3}},yt^{-\frac{%
2}{3}},~t^{-\frac{2}{3k}}U\left( \left( x-f_{0}t\right) t^{-\frac{1}{3}%
},yt^{-\frac{2}{3}}\right) ,t^{-\frac{2+k}{3k}}V\left( \left(
x-f_{0}t\right) t^{-\frac{1}{3}},yt^{-\frac{2}{3}}\right) $ \\ 
$\mathbf{X}_{5}$ & $\left( x-f_{0}t\right) t^{-\frac{1}{3}},yt^{-\frac{2}{3}%
},~-\frac{2}{3\sigma }\ln t+U\left( \left( x-f_{0}t\right) t^{-\frac{1}{3}%
},yt^{-\frac{2}{3}}\right) ,t^{-\frac{1}{3}}V\left( \left( x-f_{0}t\right)
t^{-\frac{1}{3}},yt^{-\frac{2}{3}}\right) $ \\ \hline\hline
\end{tabular}%
\label{tab4}%
\end{table}%

\subsection{Similarity solutions}

We continue by applying some of the Lie invariants presented in Table \ref%
{tab4} in order to determine similarity solutions for the 2+1 nonlinear
generalized KP equation.

\subsubsection{Static solution}

The application of the Lie symmetry vector $X_{1}$, leads to the
time-independent equation%
\begin{eqnarray}
f\left( u\right) u_{x}+u_{xxx}+\varepsilon v_{y} &=&0,  \label{so.01} \\
v_{x}-u_{y} &=&0,  \label{so.02}
\end{eqnarray}%
where $u=u\left( x,y\right) $ and $v=v\left( x,y\right) $; that is, the
solution which will be determined will be a static solution.

For arbitrary function $f\left( u\right) $ the latter equation admits the
Lie symmetry vectors $X_{2},X_{3}$ and $X_{v}=\partial _{v}$. The latter
vector fields are reduced symmetries while $X_{v}$ is the static symmetry
vector $X_{\beta }$. \ Additional symmetry vectors exist when $f\left(
u\right) =u^{k}$ and $f\left( u\right) =e^{\sigma u}$. The additional Lie
symmetries are the $X_{5}$ and $\bar{X}_{5}$ vector fields for $f_{0}=0$,
respectively. We remark that for $f_{0}\neq 0$ there are not additional Lie
point symmetries, that is because the vector fields $X_{5}$ and $\bar{X}_{5}$
become nonlocal symmetries.

Further, reduction of the system (\ref{so.01}), (\ref{so.02}) with the
application of the lie symmetry $X_{2}$ leads to the system $\varepsilon
v_{y}=0,~u_{y}=0$ with the trivial solution $v=v_{0}$ and $u=u_{0}$. On the
other hand, reduction with the use of the symmetry vector $X_{3}$ leads to
the third-order nonlinear ODE%
\begin{equation}
f\left( u\right) u_{x}+u_{xxx}=0,  \label{so.03}
\end{equation}%
where $~v=v_{0}$. Equation (\ref{so.03}) can be integrated as follows%
\begin{equation}
u_{xx}+\int f\left( u\right) du=0,  \label{so.04}
\end{equation}%
The latter equation is autonomous and can easily be integrated by
quadratures. Indeed, equation (\ref{so.04}) becomes $\frac{1}{2}%
u_{x}^{2}+\Phi \left( u\right) =0,~$where we have replaced $\int f\left(
u\right) du=\Phi _{,u}$ ;~\thinspace that is,%
\begin{equation}
\int \frac{du}{\sqrt{2\Phi \left( u\right) }}=dx.
\end{equation}

As far as the classification problem for equation (\ref{so.04}) is
concerned, that it is well-known and was performed by Sophus Lie more than a
century ago \cite{lie1}.

In particular there are four different families of potentials. (A) For
arbitrary function $F\left( u\right) $ equation (\ref{so.04}) admits the
symmetry vector $\partial _{x}$. (B) When $F\left( u\right) =\left( a+\beta
u\right) ^{n}$ or $F\left( u\right) =e^{\gamma u},~n\neq 0,1,-3$ equation (%
\ref{so.04}) admits two Lie point symmetries. Specifically the admitted Lie
point symmetries constitute the $A_{2}$ Lie algebra in the Mubarakzyanov
classification scheme. (C) Furthermore, when $F\left( u\right) =\frac{1}{%
\left( u+c\right) ^{3}}$ or $F\left( u\right) =\alpha \left( u+c\right) +%
\frac{1}{\left( u+c\right) ^{3}}$, equation (\ref{so.04}) describes the
Ermakov-Pinney equation and it is invariant under the elements of the $%
SL\left( 3,R\right) $ Lie algebra. Finally, (D) when $F\left( u\right) $ is
linear, equation (\ref{so.04}) is maximally symmetric and admits eight Lie
point symmetries. However, that case is not the subject of study of this
analysis. We note that in the case (B) the additional symmetry is a reduced
symmetry and it is described by the vector fields $X_{5}$ and $\bar{X}_{5}$.

Reduction with the Lie symmetry $\left\{ X_{2}-\gamma X_{3}\right\} $ leads
to the system%
\begin{eqnarray}
f\left( u\right) u_{z}+u_{zzz}+\varepsilon v_{z} &=&0, \\
v_{z}-u_{z} &=&0,
\end{eqnarray}%
where $z=y+cx$. The latter system is reduced in the form of equation (\ref%
{so.03}).

\subsubsection{Travel-wave solutions}

The application of the Lie point symmetries $\left\{ \mathbf{X}_{1}+\gamma 
\mathbf{X}_{2}\right\} ,~\left\{ \mathbf{X}_{1}+\gamma \mathbf{X}%
_{3}\right\} $ and $\left\{ \mathbf{X}_{1}+\gamma \mathbf{X}_{2}+\delta 
\mathbf{X}_{3}\right\} $ provides travel-wave solutions in the directions of 
$x,$ $y$ or in the line $\left\{ \gamma x+\delta y=0\right\} $.

Consider reduction of the original system with the symmetry vector $\left\{ 
\mathbf{X}_{1}+\gamma \mathbf{X}_{2}\right\} $, then it follows%
\begin{eqnarray}
\left( f\left( u\right) -\gamma \right) u_{z}+u_{zzz}+\varepsilon v_{y} &=&0,
\label{so.05} \\
v_{z}-u_{y} &=&0,  \label{so.06}
\end{eqnarray}%
where $z=x-\gamma t$. The latter system is in the form of the static system (%
\ref{so.01}), (\ref{so.02}), where someone replaces $f\left( u\right)
\rightarrow f\left( u\right) -\gamma $ and $x\rightarrow z.$ Hence the above
analysis is also applied and in that case

The same results follow and for the rest of the reductions which provide
travel-wave solutions; therefore we omit the presentation of the rest
reductions which lead to travel-wave solutions.

\section{Conclusions}

\label{sec5}

In this work, we considered a generalization of the 2+1 KP equation which
has been used for the study of weakly nonlinear waves. The generalized KP
equation depends on an unknown function $f\left( u\right) $ which we assumed
that it is constrained by the Lie symmetry conditions.

For an arbitrary function $f\left( u\right) $, the generalized KP equation
is invariant under the action of a four-dimensional Lie algebra, the $%
A_{4,3} $ Lie algebra, plus a vector field which provides the infinity
number of trivial solutions for the differential equation.

For two exact forms of $f\left( u\right) ,$ namely $f\left( u\right)
=u^{k}+f_{0}$ and $f\left( u\right) =e^{\sigma u}+f_{0}$, the generalized KP
equation admits from one additional Lie point symmetry, such that the finite
Lie algebra to be the $A_{5,37}$ and $\left\{ A_{5,37}\otimes
_{s}A_{1}\right\} $ respectively. We see that for $f\left( u\right)
=e^{\sigma u}+f_{0}$ the finite Lie algebra is of sixth dimension. However,
in both cases there exists the Lie point symmetry which provides the finite
number of trivial solutions $u=u_{0}$ and $v=v\left( t\right) $. An
important observation is that for the two different functions $f\left(
u\right) $ the two generalized KP equations has a common subalgebra, namely $%
A_{5,37}~$\ which means that they share a common reduction process, more
general than that for arbitrary function $f\left( u\right) $.

For all the different cases of $f\left( u\right) $ we derived the
one-dimensional optimal system and we calculated all the possible similarity
transformations which can be applied to reduce the differential equation. We
demonstrated our results by applying the similarity transformations to
determine analytic solutions which are static or travel-waves. Surprisingly,
we determined that for both types of solutions and after a further reduction
we end up with a similar second-order ordinary differential equation, of the
form%
\begin{equation}
X\left( \zeta \right) _{\zeta \zeta }+V\left( X\left( \zeta \right) \right)
=0,
\end{equation}%
which can be solved by quadratures.

Therefore, we conclude that the generalized 2+1 KP equation can be reduced
to a classical Newtonian system, with a central force. That is an important
result since we can see the dynamics of nonlinear waves reduce to that of
classical system under the proper frame, that is, a proper similarity
transformation. In a future work we plan to investigate in details the
physical applications of these solutions.

\bigskip

{\large {\textbf{Acknowledgements}}} \newline
The author wishes to thank Dr. Alex Giacomini and the Universidad Austral de
Chile (UACh) for the hospitality provided while this work was performed. 
\newline

\appendix

\section{Lie symmetries}

\label{ap1}

Consider the system of differential equations $%
H(x^{i},u^{A},u_{,i}^{A},u_{,ij}^{A})\equiv 0$ where $x^{i}$ denotes the
independent variables and $u^{A}$ are the dependent variables.

Under the action of the one-parameter point infinitesimal transformation 
\begin{align}
\bar{x}^{i}& =x^{i}+\varepsilon \xi ^{i}(x^{k},u^{B})~,  \label{pr.01} \\
\bar{u}^{A}& =\bar{u}^{A}+\varepsilon \eta ^{A}(x^{k},u^{B})~,  \label{pr.02}
\end{align}%
with infinitesimal generator 
\begin{equation}
\mathbf{X}=\xi ^{i}(x^{k},u^{B})\partial _{x^{i}}+\eta
^{A}(x^{k},u^{B})\partial _{u^{A}}~.  \label{pr.03}
\end{equation}%
the system of differential equations~$H(x^{i},u^{A},u_{,i}^{A},u_{,ij}^{A})$
is invariant if and only if 
\begin{equation}
\lim_{\varepsilon \rightarrow 0}\frac{\bar{H}^{A}\left( \bar{x}^{i},\bar{u}%
^{A},...;\varepsilon \right) -H^{A}\left( \bar{x}^{i},u^{A},...\right) }{%
\varepsilon }=0,
\end{equation}%
or equivalently%
\begin{equation}
\mathcal{L}_{X}\left( H\right) =0,  \label{ls.05a}
\end{equation}%
where~$\mathcal{L}$ describes the Lie derivative with respect to the vector
field $X^{\left[ n\right] }.$ Vector field $X^{\left[ n\right] }$ $n$%
th-extension of $X~$in the jet space $\left\{
x^{i},u^{A},u_{,i}^{A},u_{,ij}^{A}\right\} $ is given by the following
expression 
\begin{equation}
X^{\left[ n\right] }=X+\eta ^{\left[ 1\right] }\partial
_{u_{i}^{A}}+...+\eta ^{\left[ n\right] }\partial
_{u_{i_{i}i_{j}...i_{n}}^{A}},  \label{ls.06}
\end{equation}%
where~$\eta ^{\left[ n\right] }$ is defined as 
\begin{equation}
\eta ^{\left[ n\right] }=D_{i}\eta ^{\left[ n-1\right]
}-u_{i_{1}i_{2}...i_{n-1}}D_{i}\left( \frac{\partial \bar{x}^{j}}{\partial
\varepsilon }\right) ~,~i\succeq 1~,~\eta ^{\left[ 0\right] }=\left( \frac{%
\partial \bar{\Phi}^{A}}{\partial \varepsilon }\right) .  \label{de.08}
\end{equation}

If condition (\ref{ls.05a}) is true, then the generator $\mathbf{X}$ of the
infinitesimal transformation (\ref{pr.01})-(\ref{pr.02}) is called a Lie
point symmetry of the system of differential equations $%
H(x^{i},u^{A},u_{,i}^{A},u_{,ij}^{A})\,\,.$

The Lie invariants which correspond to a given Lie point symmetries $\mathbf{%
X}$ are found by solving the following Lagrange system%
\begin{equation}
\frac{dx^{i}}{\xi ^{i}}=\frac{du^{A}}{\eta ^{A}}=\frac{du_{i}^{A}}{\eta _{%
\left[ i\right] }^{A}}=\frac{du_{ij}^{A}}{\eta _{\left[ ij\right] }^{A}}=...%
\frac{du_{ij...j_{n}}^{A}}{\eta ^{\left[ n\right] }}
\end{equation}%
The characteristic functions $W^{\left[ 0\right] }\left( x^{k},u\right) ,~W^{%
\left[ 1\right] }\left( x^{k},u,u_{i}\right) $~and $W^{\left[ 2\right]
}\left( x^{k},u,u_{,i},u_{ij}\right) ~$which solve the latter Lagrange
system are called the $n-th$ invariants of the Lie symmetry vector $\mathbf{X%
}$.~

\section{One-dimensional optimal system}

\label{ap2}

Let assume the $n$-dimensional Lie algebra $G_{n},$ with elements $%
X_{1},~X_{2},~...~X_{n}$. We shall say that the two generic vector fields 
\begin{equation}
Z=\dsum\limits_{i=1}^{n}a_{i}X_{i}~,~W=\dsum\limits_{i=1}^{n}b_{i}X_{i}~,~%
\text{\ }a_{i},~b_{i}\text{ are constants.}  \label{sw.04}
\end{equation}%
are equivalent if and only if under the action of the Adjoint representation
it holds, 
\begin{equation}
\mathbf{W}=\dprod\limits_{j=i}^{n}Ad\left( \exp \left( \varepsilon
_{i}X_{i}\right) \right) \mathbf{Z}  \label{sw.05}
\end{equation}%
or%
\begin{equation}
W=cZ~,~c=const,  \label{sw.06}
\end{equation}%
where the Adjoint operator is defined as%
\begin{equation}
Ad\left( \exp \left( \varepsilon X_{i}\right) \right)
X_{j}=X_{j}-\varepsilon \left[ X_{i},X_{j}\right] +\frac{1}{2}\varepsilon
^{2}\left[ X_{i},\left[ X_{i},X_{j}\right] \right] +....  \label{sw.07}
\end{equation}

Hence, in order to perform a complete classification for the similarity
solutions of a given differential equation we should determine all the
one-dimensional indepedent symmetry vectors of the Lie algebra $G_{n}$. The
one-dimensional independent symmetry vectors form the so-called
one-dimensional optimal system \cite{olver}.

\section{The 3+1 nonlinear generalized\ Kadomtsev-Petviashvili equation}

\label{ap3}

The 3+1 nonlinear generalized KP equation \cite{eq1} is defined as%
\begin{eqnarray}
u_{t}+f\left( u\right) u_{x}+u_{xxx}+\alpha v_{y}+\beta w_{z} &=&0, \\
v_{x}-u_{y} &=&0, \\
w_{x}-u_{z} &=&0,
\end{eqnarray}%
or equivalently%
\begin{equation}
\left( u_{t}+f\left( u\right) u_{x}+u_{xxx}\right) _{x}+\alpha u_{yy}+\beta
u_{zz}=0,
\end{equation}%
where $u=u\left( t,x,y,z\right) ,~v=v\left( t,x,y,z\right) ,~w=w\left(
t,x,y,z\right) $ and constants $\alpha $ and~$\beta $ measures the
transverse dispersion effects and are normalized to $\pm 1$.

For the 3+1 generalized KP equation and for the arbitrary function $f\left(
u\right) $ the admitted Lie point symmetries are%
\begin{eqnarray*}
Y_{1} &=&\partial _{t}~,~Y_{2}=\partial _{x}~,Y_{3}=\partial
_{y}~,~Y_{4}=\partial _{z}~,~Y_{5}=2\alpha t\partial _{y}-y\partial
_{x}+u\partial _{v}~, \\
Y_{6} &=&2\beta t\partial _{y}-z\partial _{x}+u\partial _{w}~,~Y_{7}=\beta
y\partial _{z}-\alpha z\partial _{y}+\alpha v\partial _{w}-\beta w\partial
_{v}~, \\
Y_{\infty } &=&\phi _{1}\left( t,y,z\right) \partial _{v}+\phi _{2}\left(
t,y,z\right) \partial _{w}\,~\text{\ where }\alpha \phi _{1y}+\beta \phi
_{2z}=0.
\end{eqnarray*}

When $f\left( u\right) =u^{k}+f_{0}$ the additional Lie point symmetry is%
\begin{equation*}
Y_{8}=k\left( 3t\partial _{t}+\left( x+2f_{0}t\right) \partial
_{x}+y\partial _{y}+z\partial _{z}\right) -2u\partial _{u}+\left( k+2\right)
\left( v\partial _{v}+w\partial _{w}\right) ,
\end{equation*}%
while when $f\left( u\right) =e^{\sigma u}+f_{0}$ the extra Lie point
symmetry of the 3+1 generalized KP equation is 
\begin{equation*}
\bar{Y}_{8}=\sigma \left( 3t\partial _{t}+\left( x+2f_{0}t\right) \partial
_{x}+y\partial _{y}+z\partial _{z}+v\partial _{v}+w\partial _{w}\right)
-2\partial _{u}.
\end{equation*}

\end{document}